\documentclass[pra,aps,twocolumn,epsfig,superscriptaddress,showpacs]{revtex4}
\usepackage{amsmath}
\usepackage{amssymb}
\usepackage[dvips]{graphicx}
\usepackage{dcolumn}
\usepackage{bm}
\usepackage[colorlinks=false,dvipdfm]{hyperref} 
\usepackage{setspace}
\usepackage{subfigure}
\usepackage{bm}

\begin{document}
\author{Rui-Juan Gu}
\affiliation{School of Mathematical Sciences, Capital Normal
University, Beijing 100048, People's Republic of China}
\author{Fu-Lin Zhang}\email{flzhang@tju.edu.cn}
\affiliation{Physics Department, School of Science, Tianjin
University, Tianjin 300072, China}
\author{Shao-Ming Fei}
\affiliation{School of Mathematical Sciences, Capital Normal
University, Beijing 100048, People's Republic of China}
\author{Jing-Ling Chen}
\affiliation{Theoretical Physics Division, Chern Institute of
Mathematics, Nankai University, Tianjin 300071, People's Republic of
China}

\date{\today}

\title{Alternative Decomposition of Two-Qutrit Pure States and
Its Relation with Entanglement Invariants\footnote{International
Journal of Quantum Information Vol. 9, No. 6 (2011)
1499-1509\\http://dx.doi.org/10.1142/S0219749911008040} }

\begin{abstract}
Based on maximally entangled states in the full- and sub-spaces of
two qutrits, we present an alternative decomposition of two-qutrit
pure states in a form
$|\Psi\rangle=\frac{p_{1}}{\sqrt{3}}(|00\rangle+|11\rangle+|22\rangle)
+\frac{p_{2}}{\sqrt{2}}(|01\rangle+|12\rangle)+
p_{3}e^{i\theta}|02\rangle$. Similar to the Schmidt decomposition,
all two-qutrit pure states can be transformed into the alternative
decomposition under local unitary transformations, and the parameter
$p_1$ is shown to be an entanglement invariant.\\
\end{abstract}

\pacs{03.67.-a, 03.67.Mn, 03.65.-w}

 \keywords{Decomposition of quantum states; Two qutrits;
Quantum information; Entanglement invariants}

\maketitle

\section{Introduction}
Decomposition of quantum states is an interesting topic in quantum
information theory \cite{Englert,Carteret,Acin}. Given an arbitrary
bipartite state, it is well-known that the Schmidt decomposition is
always applicable \cite{Book}. For instance, under local unitary
transformations any two-qubit state $|\Psi\rangle=\sum_{i,j=0}^{1}
a_{ij}|i\rangle_A |j\rangle_B$ can be transformed into its
Schmidt-form as $|\Psi'\rangle=U_A \otimes U_B |\Psi\rangle =
\kappa_1 |00\rangle + \kappa_2 |11 \rangle$.

Besides the Schmidt decomposition, other decompositions are
possible. For example, in 2001, Abouraddy \emph{et al.} have
proposed an alternative decomposition for two-qubit pure states
based on the maximally entangled state \cite{Abouraddy}:
\begin{eqnarray}\label{2qubit-state}
|\Psi\rangle=p_1 |\Psi\rangle_e + p_2 \; e^{i \varphi}
|\Psi\rangle_f,
\end{eqnarray}
where $p_1 \ge 0$, $p_2=\sqrt{1-p_1^2}$, $|\Psi\rangle_e$ is the
two-qubit maximally entangled state, and $|\Psi\rangle_f$ is a
factorizable state orthogonal to $|\Psi\rangle_e$. They showed that
such a decomposition \emph{always exists and is not unique, but the
parameter $p_1$ is unique}. In comparison to the Schmidt
decomposition, the merit of the new kind of decomposition is that
the parameter $p_1$ has a definite physical significance as the
degree of entanglement of two qubits. In this work, we would like
to generalize the alternative decomposition to a two-qutrit system
based on the maximally entangled states in the full- and sub-spaces.
To our knowledge, such a generalization has not been reported in the
literature.

This paper is organized as follows: In section II, we make a brief
review for the previous result of Abouraddy \emph{et al.}, but from
a different viewpoint of entanglement invariants. In section III, we
present a \emph{Theorem} on the alternative decomposition of
two-qutrit pure states, and also show its relation with the
entanglement invariants. Conclusion and discussion are made in the
last section.


\section{Brief Review of Entanglement Invariants and Previous Result of Abouraddy \emph{et al.}}

Let us consider a general pure state of two $d$-dimensional quantum
systems (two qudits), which takes of the following form:
\begin{eqnarray}\label{2qudit-state}
|\Psi\rangle_{AB}=\sum_{i,j=0}^{d-1} a_{ij}\; |i\rangle_A
|j\rangle_B,
\end{eqnarray}
where $|i\rangle_A$ and $|j\rangle_B$ are the orthonormal bases of
the Hilbert spaces A and B respectively, and $a_{ij}$'s are complex
numbers satisfying the normalization condition $\sum_{i,j=0}^{d-1}
|a_{ij}|^2=1$.

Let $\mathcal {A}$ denote the matrix whose matrix elements are given
by $(\mathcal {A})_{ij}=a_{ij}$. It has been shown that the
following quantities are entanglement invariants under local unitary
transformations \cite{Fei}:
\begin{eqnarray}\label{2qudit-invariant1}
I_{n}={\rm Tr}[(\mathcal {A} {\mathcal {A}}^{\dag})^{n+1}], \; n=0,
1, ..., d-1.
\end{eqnarray}
Denote $\rho_{AB}=|\Psi\rangle_{AB} {_{AB}}\langle \Psi|$, since the
reduced density matrices $\rho_A={\rm Tr}_B[\rho_{AB}]=\mathcal {A}
{\mathcal {A}}^{\dag}$, $\rho_B={\rm Tr}_A[\rho_{AB}]= {\mathcal
{A}}^{\dag}\mathcal {A}$, thus Eq. (\ref{2qudit-invariant1}) can be
also expressed as
\begin{eqnarray}\label{2qudit-invariant2}
I_{n}={\rm Tr}[\rho_A^{n+1}]={\rm Tr}[\rho_B^{n+1}], \; n=0, 1, ...,
d-1.
\end{eqnarray}
For $n=0$, one easily has $I_{0}=1$, which is nothing but the
normalization condition of the reduced density matrix of $\rho_A$ or
$\rho_B$. Therefore, for a two-qudit system, there are only $(d-1)$
nontrivial entanglement invariants.

After performing an appropriate local unitary transformation, one
may transform the general state $|\Psi\rangle_{AB}$ into its
Schmidt-form as
\begin{eqnarray}\label{SD-2qudit}
|\Psi\rangle_{2-qudit}=\kappa_1 |00\rangle+\kappa_2
|11\rangle+\cdots+\kappa_{d} |d-1, d-1\rangle,
\end{eqnarray}
where $\kappa_j$'s $(j=1, 2, ..., d)$ are the Schmidt coefficients,
which satisfy the normalization condition:
$\sum_{j=1}^{d}|\kappa_j|^2$=1. In the Schmidt representation, it is
easy to obtain the entanglement invariants as
\begin{eqnarray}\label{2qudit-invariant3}
I_{n}={\rm Tr}[\rho_A^{n+1}]={\rm Tr}[\rho_B^{n+1}]=\sum_{j=1}^{d}
|\kappa_j|^{2(n+1)}.
\end{eqnarray}

Now, the previous result of Abouraddy \emph{et al.} can be
re-expressed as the following theorem:

\emph{Theorem 1.}  Under local unitary transformations any two-qubit
state can be always transformed into an alternative decomposition as
\begin{eqnarray}\label{2qubit-decom}
&&|\Psi\rangle_{AB} = p_1 \frac{1}{\sqrt{2}}(|00\rangle + |11
\rangle)
+ p_2 |01\rangle, \\
&& p_1 \ge 0, \;\; p_2=\sqrt{1-p_1^2},\nonumber
\end{eqnarray}
where $p_1^4 = 2(I_0-I_1)=4 \kappa_1^2 \kappa_2^2= 4 {\rm
Det}[\rho_A]$ is unique and is an entanglement invariant under the
local unitary transformations.

By comparing Eq. (\ref{2qubit-state}) and Eq. (\ref{2qubit-decom}),
one notes that we have chosen the maximally entangled state of
two-qubit as $|\Psi\rangle_e=\frac{1}{\sqrt{2}}(|00\rangle + |11
\rangle)$ and the factorizable state as $|\Psi\rangle_f=|01\rangle$.
Moreover, the phase factor $e^{i \varphi}$ in Eq.
(\ref{2qubit-state}) can be eliminated further by a suitable
$U(1)\otimes U(1)$ transformation. Therefore the decomposition in
Eq. (\ref{2qubit-decom}) is unique for the pure states of a
two-qubit system.

The standard way to prove \emph{Theorem 1} is owing to the local
unitary transformations, which has been actually given in Ref.
\cite{Abouraddy}, namely, by acting the appropriate local unitary
transformations $U_A \otimes U_B$ on an arbitrary two-qubit pure
state $|\Psi\rangle=\sum_{i,j=0}^{1} a_{ij}|i\rangle_A |j\rangle_B$,
then one obtains the decomposition (\ref{2qubit-decom}). However,
there is another equivalent way to prove \emph{Theorem 1}, which is
due to the entanglement invariants. Now we use the new approach to
prove \emph{Theorem 1}, the same approach will be used to prove the
corresponding Theorem for the two-qutrit case.

\emph{Proof.} On one hand, for the two-qubit state in the
Schmidt-form
\begin{eqnarray}\label{SD-2qubit}
|\Psi\rangle_{2-qubit}=\kappa_1 |00\rangle+\kappa_2 |11\rangle,
\end{eqnarray}
one has the entanglement invariants as
\begin{eqnarray}\label{2qubit-Schmidt}
I_{0}[\vec{\kappa}]&=& |\kappa_1|^2+|\kappa_2|^2=1, \nonumber\\
I_{1}[\vec{\kappa}]&=& |\kappa_1|^4+|\kappa_2|^4,
\end{eqnarray}
here $\vec{\kappa}=(\kappa_1, \kappa_2)$, $I_{n}[\vec{\kappa}]$
means that $I_{n}$ is expressed by the parameters $\kappa_1$ and
$\kappa_2$.

On the other hand, for the two-qubit state in the alternative
decomposition as in Eq.  (\ref{2qubit-decom}), one has the matrices
\begin{eqnarray}\label{2qubit-matrix-A}
\mathcal {A}=
\left(%
    \begin{array}{cc}
      \frac{p_{1}}{\sqrt{2}} & p_2 \\
      0 & \frac{p_{1}}{\sqrt{2}}
    \end{array}%
    \right), \;\;
{\mathcal {A}}^\dag=
\left(%
    \begin{array}{cc}
      \frac{p_{1}}{\sqrt{2}} & 0 \\
      p_2 & \frac{p_{1}}{\sqrt{2}}
    \end{array}%
    \right).
\end{eqnarray}
Thus the corresponding entanglement invariants reads
\begin{eqnarray}\label{2qubit-alter}
I_{0}[\vec{p}]&=& p_1^2+p_2^2=1, \nonumber\\
I_{1}[\vec{p}]&=& 1- \frac{p_1^4}{2},
\end{eqnarray}
here $\vec{p}=(p_1, p_2)$, $I_{n}[\vec{p}]$ means that $I_{n}$ is
expressed by the parameters $p_1$ and $p_2$.


Because an arbitrary two-qubit state can be transformed into the
Schmidt decomposition under the local unitary transformation, if one
can prove that for any given $\kappa_1$ and $\kappa_2$, there always
exists $\vec{p}$ satisfying $I_{n}[\vec{p}]=I_{n}[\vec{\kappa}]$,
($n=0, 1$), then it implies that an arbitrary two-qubit state can be
transformed into the alternative decomposition as shown in Eq.
(\ref{2qubit-decom}) under the local unitary transformation. Since
$I_{0}[\vec{p}]=I_{0}[\vec{\kappa}]=1$ is the normalization
condition, one only need to study
$I_{1}[\vec{p}]=I_{1}[\vec{\kappa}]$, this yields the following
solution:
\begin{eqnarray}\label{2qubit-p1}
p_1^4=2(I_0-I_1)=4 \kappa_1^2 \kappa_2^2 \in [0,1],
\end{eqnarray}
which means that an arbitrary two-qubit state can be transformed
into the alternative decomposition (\ref{2qubit-decom}) under the
local unitary transformation if relation (\ref{2qubit-p1}) is
satisfied. This ends the proof.

By the way, it is easy to show that the determinants of matrices
$\mathcal {A}$ and ${\mathcal {A}}^\dag$ are
\begin{eqnarray}\label{det-A}
{\rm Det}[\mathcal {A}]={\rm Det}[{\mathcal
{A}}^\dag]=\frac{p_1^2}{2},
\end{eqnarray}
therefore one has
\begin{eqnarray}\label{p1-det-A}
p_1^4 =4 {\rm Det}[\mathcal {A}]\; {\rm Det}[{\mathcal {A}}^\dag]=4
{\rm Det}[\mathcal {A}{\mathcal {A}}^\dag ]= 4 {\rm Det}[\rho_A].
\end{eqnarray}
One will find later that such a similar relation holds for the any
two-qudit system.

\section{Entanglement Invariants of Two-Qutrit and the Alternative Decomposition}

Under local unitary transformations an arbitrary two-qutrit pure
state can be transformed into its Schmidt-form as
\begin{eqnarray}\label{SD-2qutrit}
|\Psi\rangle_{2-qutrit}=\kappa_1 |00\rangle+\kappa_2
|11\rangle+\kappa_3 |22\rangle,
\end{eqnarray}
one has the entanglement invariants as
\begin{eqnarray}\label{2qutrit-Schmidt}
I_{0}[\vec{\kappa}]&=& |\kappa_1|^2+|\kappa_2|^2+|\kappa_3|^2=1, \nonumber\\
I_{1}[\vec{\kappa}]&=& |\kappa_1|^4+|\kappa_2|^4+|\kappa_3|^4, \nonumber\\
I_{2}[\vec{\kappa}]&=& |\kappa_1|^6+|\kappa_2|^6+|\kappa_3|^6,
\end{eqnarray}
here $\vec{\kappa}=(\kappa_1, \kappa_2,  \kappa_3)$, and
$I_{0}[\vec{\kappa}]=1$ is trivial as the normalization condition of
a quantum state.

By expanding
$(I_{0}[\vec{\kappa}])^3=(|\kappa_1|^2+|\kappa_2|^2+|\kappa_3|^2)^3$,
one may get an interesting and useful relation:
\begin{eqnarray}\label{intere-1}
I_{2}[\vec{\kappa}]-\frac{3}{2}I_{1}[\vec{\kappa}]=-\frac{1}{2}
I_{0}[\vec{\kappa}] + 3 \mathcal {K},
\end{eqnarray}
with
\begin{eqnarray}\label{intere-1a}
\mathcal {K}= \kappa_{1}^{2}\kappa_{2}^{2}\kappa_{3}^{2}.
\end{eqnarray}
Since $I_{1}[\vec{\kappa}]$ and $I_{2}[\vec{\kappa}]$ are
entanglement invariants, thus $\mathcal {K}$ is an entanglement
invariant under local unitary transformation. $\mathcal
{K}\in[0,\frac{1}{27}],$ $\mathcal {K}$ reaches its maximum value
$\frac{1}{27}$ when
$\kappa_{1}^{2}=\kappa_{2}^{2}=\kappa_{3}^{2}=\frac{1}{3}$. We shall
use such a useful relation to prove the \emph{Theorem 2} in this
section.

Actually, the entanglement property of a two-qutrit system is
completely characterized by two entanglement invariants
$I_{1}[\vec{\kappa}]$ and $I_{2}[\vec{\kappa}]$, or equivalently,
\begin{eqnarray}\label{invariant}
&& I'_1[\vec{\kappa}]=\frac{3}{2}(1-I_1[\vec{\kappa}]),\nonumber\\
&&I'_2[\vec{\kappa}]=\frac{9}{8}(1-I_2[\vec{\kappa}]),
\end{eqnarray}
where the normalized entanglement invariants $I'_1, I'_2 \in [0,
1]$.


\begin{figure}\label{fig1}
\includegraphics[width=75mm]{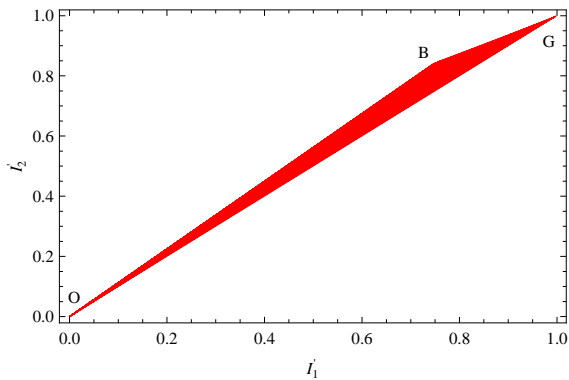}\\
 \caption{(Color online) In the $I'_1-I'_2 $ coordinate,
the factorizable state,
 such as $|00\rangle$, locates at the origin $O=(0, 0)$; the maximally entangled
 state (or say the GHZ state) in the full-space
of two-qutrit locates at the point $G=(1,
 1)$, which is the point farthest from the origin; and the maximally entangled
 state in the sub-space of
two-qutrit, such as  $\frac{1}{\sqrt{2}} ( |00\rangle+ |11\rangle)$,
locates at the point $B=(\frac{3}{4}, \frac{27}{32})$.}
\end{figure}

In Fig.1, we have plots points $(I'_1, I'_2)$ for the two-qutrit
state $|\psi\rangle_{2-qutrit}=\kappa_1 |00\rangle+\kappa_2
|11\rangle+\kappa_{3} |22\rangle$ by randomly taking $10^7$ values
of $\kappa_1$, $\kappa_2$, and $\kappa_3$, see the red region of
figure, whose contour lines form a curved triangle $\Delta OBG$. In
the $I'_1-I'_2 $ coordinate, one may observe that there are three
special points: the first point is the origin $O=(0, 0)$, which
corresponds to the factorizable states, such as $|00\rangle$; the
second is the point $G=(1, 1)$, which corresponds to the maximally
entangled state (or say the GHZ state) in the full-space of
two-qutrit, such as $|\psi\rangle_{GHZ}=\frac{1}{\sqrt{3}} (
|00\rangle+ |11\rangle+|22\rangle)$; and the third is the point
$B=(\frac{3}{4}, \frac{27}{32})$, which corresponds to the entangled
state in the sub-space of two-qutrit, such as $\frac{1}{\sqrt{2}} (
|00\rangle+ |11\rangle)$.

Inspired by the success of \emph{Theorem 1}, we suggest the
following decomposition for two-qutrit pure states:
\begin{eqnarray}\label{2qutrit-decom-old}
&&|\Psi\rangle_{AB} = p_1 |\psi_1\rangle+ p_2 |\psi_2\rangle+p_3
|\psi_3\rangle,\\
&& p_1 \ge 0, \;\;  p_2 \ge 0, \;\;
p_3=\sqrt{1-p_1^2-p_2^2},\nonumber
\end{eqnarray}
and
\begin{eqnarray}\label{2qutrit-psi123}
|\psi_1\rangle &=& \frac{1}{\sqrt{3}} ( e^{i \theta_1}|00\rangle+
e^{i \theta_2} |11\rangle+e^{i \theta_3}|22\rangle), \nonumber\\
|\psi_2\rangle &=& \frac{1}{\sqrt{2}} ( e^{i \theta_4}|01\rangle+
e^{i \theta_5} |12\rangle), \nonumber\\
|\psi_3\rangle &=& e^{i \theta_6}|02\rangle.
\end{eqnarray}
Here $|\psi_1\rangle$ is the maximally entangled state (or say the
GHZ state) in the full-space of two-qutrit spanned by $\{|00\rangle,
|11\rangle, |22\rangle\}$, $|\psi_2\rangle$ is the maximally
entangled state in the sub-space of two-qutrit spanned by
$\{|01\rangle, |12\rangle\}$, and $|\psi_3\rangle$ is the
factorizable state, they are mutually orthogonal, i.e., $\langle
\psi_i| \psi_j \rangle=\delta_{ij}$. $\theta_j$'s $(j=1, 2, ..., 6)$
are some phase factors. However, five phases can be eliminated by
the transformation $U_a \otimes U_b$, with $U_a= \sum_{j=0}^{2} e^{i
\phi_a^j} |j\rangle \langle j|$ and $U_b= \sum_{j=0}^{2} e^{i
\phi_b^j} |j\rangle \langle j|$, thus there is only one phase factor
is survival. In general, one may select the phase factor involved in
$|\psi_3\rangle$ is not zero. Consequently, one arrives at the
alternative decomposition of two-qutrit pure states as follows:
$|\Psi\rangle_{AB} = p_1 \frac{1}{\sqrt{3}} ( |00\rangle+
|11\rangle+|22\rangle)+ p_2 \frac{1}{\sqrt{2}} ( |01\rangle+
|12\rangle)+p_3 e^{i \theta}|02\rangle$.

Our main result is the following Theorem.

\emph{Theorem 2.}  Under local unitary transformations any
two-qutrit state can be always transformed into an alternative
decomposition as
\begin{eqnarray}\label{2qutrit-decom-theorem}
|\Psi\rangle_{AB} &=& p_1 \frac{1}{\sqrt{3}} ( |00\rangle+
|11\rangle+|22\rangle)+ \nonumber\\
&&p_2 \frac{1}{\sqrt{2}} ( |01\rangle+ |12\rangle)+p_3 e^{i
\theta}|02\rangle, \\&& p_1 \ge 0, \;\;  p_2 \ge 0, \;\;
p_3=\sqrt{1-p_1^2-p_2^2},\nonumber
\end{eqnarray}
where $p_1^6 = 9(I_2-\frac{3}{2}I_1+\frac{1}{2}I_0)=27 {\rm
Det}[\rho_A]$ is unique and is an entanglement invariant under the
local unitary transformations.

\emph{Proof. } Similarly, for the 2-qutrit pure quantum state in
form (\ref{2qutrit-decom-theorem}), one can write the related
matrices as
\begin{eqnarray}
\mathcal{A}=
\left(%
    \begin{array}{ccc}
      \frac{p_{1}}{\sqrt{3}} & \frac{p_{2}}{\sqrt{2}} & p_{3}e^{i\theta} \\
      0 & \frac{p_{1}}{\sqrt{3}} & \frac{p_{2}}{\sqrt{2}} \\
      0 & 0 & \frac{p_{1}}{\sqrt{3}} \\
    \end{array}%
    \right),
   \mathcal{A}^{\dag}=
\left(%
    \begin{array}{ccc}
      \frac{p_{1}}{\sqrt{3}} & 0 & 0 \\
      \frac{p_{2}}{\sqrt{2}} & \frac{p_{1}}{\sqrt{3}} & 0 \\
      p_{3}e^{-i\theta} & \frac{p_{2}}{\sqrt{2}} & \frac{p_{1}}{\sqrt{3}} \\
    \end{array}%
    \right).
\end{eqnarray}
Its entanglement invariants are obtained immediately
\begin{eqnarray}\label{2qutrit-alter}
I_{0}[\vec{p}]&=&p_{1}^{2}+p_{2}^{2}+p_{3}^{2}=1,\nonumber\\
I_{1}[\vec{p}]&=&1-\frac{2}{3}p_{1}^{2}-\frac{1}{2}p_{2}^{4}+\frac{2}{\sqrt{3}}p_{1}p_{2}^{2}p_{3}\cos\theta,\nonumber\\
I_{2}[\vec{p}]&=&1-p_{1}^{2}+\frac{p_{1}^{6}}{9}-\frac{3}{4}p_{2}^{4}+\sqrt{3}p_{1}p_{2}^{2}p_{3}\cos\theta.
\end{eqnarray}
From them, one can find the relation
\begin{eqnarray}\label{2qutrit-alter1}
I_{2}[\vec{p}]-\frac{3}{2}I_{1}[\vec{p}]=-\frac{1}{2}I_{0}[\vec{p}]+\frac{p_{1}^{6}}{9}.
\end{eqnarray}
On condition that the state in Eq. (\ref{2qutrit-decom-theorem}) is
equivalent to the one in Eq. (\ref{SD-2qutrit}) under local unitary
(LU) transformations, the parameter $p_{1}$ should satisfies
\begin{eqnarray}\label{2qutrit-alter2}
p_{1}^{6}=27 {\rm
Det}[\rho_A]=27\mathcal{K}=27\kappa_{1}^{2}\kappa_{2}^{2}\kappa_{3}^{2},
\end{eqnarray}
which always has a root in the  interval $p_{1}\in [0,1]$ for any
value of $\mathcal{K} \in [0,1/27]$. Then, the two nontrivial
entanglement invariants can be replaced by
\begin{eqnarray}
\mathcal{I}_1&=&I_1,\nonumber \\
\mathcal{I}_2&=&I_{2}-\frac{3}{2}I_{1}.
\end{eqnarray}
For a fixed value of
$\mathcal{I}_2[\vec{p}]=\mathcal{I}_2[\vec{\kappa}]=\mathcal{I}_2$,
if the range of $\mathcal{I}_1[\vec{p}]=I_1[\vec{p}]$ in Eq.
(\ref{2qutrit-alter}) is the same as the one of
$\mathcal{I}_1[\vec{\kappa}]=I_1[\vec{\kappa}]$ in Eq.
(\ref{2qutrit-Schmidt}), one can conclude there exists a pure state
in the form (\ref{2qutrit-decom-theorem}) equivalent the one
(\ref{SD-2qutrit}) with any $\vec{\kappa}$ under LU transformations.
Let us denote the minimum and maximum of $\mathcal{I}_1$ as
$\underline{\mathcal{I}}_1$ and $\overline{\mathcal{I}}_1$. Based on
the fact that the values of $\mathcal{I}_1[\vec{\kappa}]$ and
$\mathcal{I}_1[\vec{p}]$ vary continuously from their minimums to
maximums, it is only to prove
\begin{eqnarray}\label{IpIk}
\underline{\mathcal{I}}_1[\vec{\kappa}]&=&\underline{\mathcal{I}}_1[\vec{p}],
\nonumber \\
\overline{\mathcal{I}}_1[\vec{\kappa}]&=&\overline{\mathcal{I}}_1[\vec{p}],
\end{eqnarray}
for a given value of $\mathcal{I}_2$. In Appendix \ref{appa}, we
show the two relations come into existence. Since an arbitrary two-
qutrit pure state can be transformed into the form
(\ref{2qutrit-Schmidt}) under LU operation, it can always be
decomposed as Eq. (\ref{2qutrit-decom-theorem}). This ends the
proof.

\section{Conclusion and Discussion}

In conclusion, we show that all 2-qutrit pure states can be
rewritten as
$|\Psi\rangle=\frac{p_{1}}{\sqrt{3}}(|00\rangle+|11\rangle+|22\rangle)+\frac{p_{2}}{\sqrt{2}}(|01\rangle+|12\rangle)+
p_{3}e^{i\theta}|02\rangle.$ The method we have used is to verify
the invariant space is as same as achieved by expression of
Schimidt-form. The parameter $p_{1}\in[0,1]$ is unique and it is an
entanglement invariant under LU operations. The values of $p_2$ and
$\theta$ can be derived from the relations in Eq.
(\ref{2qutrit-decom-theorem}).

 In this paper, we concerns us in the
pure states of two-qutrit system. There are two natural extensions
of this issue: (\romannumeral1) to decompose the pure states in a
bipartite arbitrary-dimensional system,  (\romannumeral2) to
decompose the pure states in a multipartite system. For the case
(\romannumeral1), we can foretell a two-qudit state can be
transformed as
\begin{eqnarray}\label{2qudit-alter}
| \Psi \rangle_{2-qudit}=\sum^{d}_{M=1}p_{d-M+1} \sum^{M-1}_{m=0}
\frac{e^{i\theta ^{M}_{m}}}{\sqrt{M}} |m,d-M+m \rangle,
\end{eqnarray}
where the parameters $\theta ^{M}_{m}\in[0,2\pi]$, $p_{d-M+1} \in
[0,1]$ and $\sum^{d}_{M=1}p^2_{d-M+1}=1$. And, here
$\sum^{M-1}_{m=0} \frac{e^{i\theta ^{M}_{m}}}{\sqrt{M}} |m,d-M+m
\rangle$ is a maximally entangled state in the sub-space $\{|m,d-M+m
\rangle | m=0,...,M-1\}$, whose spacial case is shown in Eq.
(\ref{2qutrit-psi123}) for $d=3$. Under locally phase
transformations $U_a \otimes U_b= \sum_{j=0}^{d-1} e^{i \phi_a^j}
|j\rangle \langle j| \otimes \sum_{k=0}^{d-1} e^{i \phi_b^k}
|k\rangle \langle k|$, the phases $\theta ^{d}_{m}$ and $\theta
^{d-1}_{m}$ can be eliminated. We have numerically verified that the
entanglement invariants of the states (\ref{2qudit-alter}) cover the
the ones of Schmidt-form states (\ref{2qudit-state}) perfectly for
$d=4$. For (\romannumeral2), the quantum correlation or entanglement
in a multipartite state carry more nonclassical characteristics of
quantum mechanics \cite{Linden02,Svet}. Many perspectives have been
presented to attempt an understanding of the problem in recent
studies \cite{Svet,Linden02,Zhou08,Linden02W,WL1,WL2,MSPRL}. In our
subsequent investigation, we hope to give a decomposition of a
multipartite pure state, dividing it into sub-spaces which reflect
the entanglement in different levels.

\begin{acknowledgments}
FLZ is supported by NSF of China (Grant No. 11105097). JLC is
supported by National Basic Research Program (973 Program) of China
under Grant No. 2012CB921900 and NSF of China (Grant Nos. 10975075
and 11175089).
\end{acknowledgments}

\bibliography{invariant}


\appendix

\section{Equivalence of the Ranges of $\mathcal{I}_1[\vec{\kappa}]$
and $\mathcal{I}_1[\vec{p}]$}\label{appa}

\emph{a. $\underline{\mathcal{I}}_1[\vec{\kappa}]$ and
$\overline{\mathcal{I}}_1[\vec{\kappa}]$.} Firstly, for the
Schmidt-decomposed state (\ref{SD-2qutrit}),we consider the extremal
values of $\mathcal{I}_1[\vec{\kappa}]$, when
$\mathcal{I}_2[\vec{\kappa}]$ (or say $\mathcal{K}$) is fixed. From
the relations (\ref{2qutrit-Schmidt}), one can obtain
\begin{eqnarray}\label{def}
&&\mathcal{I}_{1}[\vec{\kappa}]=\mathcal{I}_{1}[\kappa_{1}^{2}]=2(-\kappa_{1}^{2}+\kappa_{1}^{4}-\frac{\mathcal
{K}}{\kappa_{1}^{2}})+1, \label{I1k1} \\
&&\kappa_{2}^{4}-(1-\kappa_{1}^{2})\kappa_{2}^{2}+\frac{\mathcal
{K}}{\kappa_{1}^{2}}=0 \label{k1k2}.
\end{eqnarray}
Then the problem is transformed to derive extremal values of
$\mathcal{I}_{1}[\vec{\kappa}]$ in Eq. (\ref{I1k1}) in the range
$\kappa_{1}\in[0,1]$, under the constraint that the values of
$\kappa_{2}$ and $\kappa_{3}$ should be legitimate. Solving the Eq.
(\ref{k1k2}), we find
\begin{eqnarray}
\kappa_{2,3}^{2}=\frac{1}{2}(1-\kappa_{1}^{2})\pm\frac{1}{2}\sqrt{(1-\kappa_{1}^{2})^{2}-\frac{4\mathcal
{K}}{\kappa_{1}^{2}}},
\end{eqnarray}
or permutation. Therefore the constraint can be explicitly expressed
as the discriminant
\begin{eqnarray}
(1-\kappa_{1}^{2})^{2}-\frac{4\mathcal {K}}{\kappa_{1}^{2}} \geq 0.
\end{eqnarray}
This leads to $\kappa^2_1 \in [t_{-},t_{+}]$, where $t_{\pm}$ are
two of the roots of the cubit equation $t^{3}-2t^{2}+t-4\mathcal
{K}=0$. They are given by
\begin{eqnarray}
t_{\pm}=\frac{2}{3}(1+\cos \frac{\phi_{1}\mp 2\pi}{3}),
\end{eqnarray}
where the angle satisfies $\cos\phi_{1}=54\mathcal {K}-1 \in[-1,1]$.
The minimal value of $\mathcal{I}_{1}[\vec{\kappa}]$ occurs when
$\kappa^2_1 = t_{-}$ and $\kappa^2_2=\kappa^2_3=(1-t_{-})/2$ or
\begin{eqnarray}\label{ptial}
\frac{\partial\mathcal{I}_{1}[\kappa^2_1]}{\partial(\kappa^2_1)}=0,\
\
\frac{\partial^2\mathcal{I}_{1}[\kappa^2_1]}{\partial(\kappa^2_1)^2}>
0.
\end{eqnarray}
Substituting the solutions of Eq. (\ref{ptial}) into Eq.
(\ref{k1k2}), one can find the result is only a permutation of the
former case, \emph{e.g.} $\kappa^2_2 = t_{-}$ and
$\kappa^2_1=\kappa^2_3=(1-t_{-})/2$. In the same way, one can
conclude that the maximal value of $\mathcal{I}_{1}[\vec{\kappa}]$
occurs when $\kappa^2_1 = t_{+}$ and
$\kappa^2_2=\kappa^2_3=(1-t_{+})/2$. Uniformly, we write the
 minimum and maximum of $\mathcal{I}_{1}[\vec{\kappa}]$ as
 $\underline{\mathcal{I}}_{1}[\vec{\kappa}]=\mathcal{I}_{1}[t_{-}]$ and
 $\overline{\mathcal{I}}_{1}[\vec{\kappa}]=\mathcal{I}_{1}[t_{+}]$ with
 \begin{eqnarray}\label{def4}
 \frac{ \mathcal{I}_{1}[t_{\pm}]-1}{2}
 =\frac{4}{9}\bigr(1+\cos\frac{\phi_{1}\mp
 2\pi}{3}\bigr)^{2}-\frac{2}{3}\bigr(1+\cos\frac{\phi_{1}\mp2\pi}{3}\bigr) \nonumber \\
 -
 \frac{3\mathcal
{K}}{2\bigr(1+\cos\frac{\phi_{1} \mp 2\pi}{3}\bigr)}.\ \ \
\end{eqnarray}


\emph{b. $\underline{\mathcal{I}}_1[\vec{p}]$ and
$\overline{\mathcal{I}}_1[\vec{p}]$.}
 For the pure states
(\ref{2qutrit-decom-theorem}), when the parameter $p_2$ or say the
entanglement invariant $\mathcal{I}_{2}[\vec{p}]$ is fixed,
$\mathcal{I}_{1}[\vec{p}]$ can be expressed as the function of $p_3$
and $\theta$
\begin{eqnarray}\label{I1p3}
\mathcal{I}_{1}[\vec{p}]
=1-\frac{2}{3}p_{1}^{2}-\frac{1}{2}(1-p_{1}^{2}-p_{3}^{2})^{2}
\nonumber\\
+\frac{2}{\sqrt{3}}p_{1}p_{3}(1-p_{1}^{2}-p_{3}^{2})\cos\theta.
\end{eqnarray}
Because $p_{1}p_{3}(1-p_{1}^{2}-p_{3}^{2})\geq 0$, the maximum value
of $\mathcal{I}_{1}[\vec{p}]$ happens at $\cos\theta=1$ and the
minimum one at $\cos\theta=-1$.

When $\cos\theta=-1$, the derivative on Eq. (\ref{I1p3})
$\partial\mathcal{I}_{1}[\vec{p}] /
\partial p_{3}=0$ leads to
\begin{eqnarray}\label{def6}
p_{3}^{3}-\sqrt{3}p_{1}p_{3}^{2}+(p_{1}^{2}-1)p_{3}+\frac{p_{1}(1-p_{1}^{2})}{\sqrt{3}}=0.
\end{eqnarray}
One of its three roots lies in $[0,\sqrt{1-p^2_1}]$ being
\begin{eqnarray}
p_{3}=\frac{1}{\sqrt{3}}(p_{1}+2\cos\frac{\phi_{2}-2\pi}{3}),
\end{eqnarray}
where $\cos\phi_{2}=p_{1}^{3}$. It corresponds to the minimal value
of  $\mathcal{I}_{1}[\vec{p}]$ as
 \begin{eqnarray}\label{def7}
 \underline{\mathcal{I}}_{1}[\vec{p}]=\frac{1}{2}+\frac{8}{9}p_{1}^{3}\cos\frac{\phi_{2}-2\pi}{3}+\frac{4}{3}\cos^{2}\frac{\phi_{2}-2\pi}{3} \nonumber \\ -
 \frac{8}{9}\cos^{4}\frac{\phi_{2}-2\pi}{3}.
\end{eqnarray}
When $\cos\theta=1$, by completely the same analysis, we obtain the
maximum
\begin{eqnarray}\label{def9}
 \overline{\mathcal{I}}_{1}[\vec{p}]=\frac{1}{2}-\frac{8}{9}p_{1}^{3}\cos\frac{\phi_{3}}{3}+\frac{4}{3}\cos^{2}\frac{\phi_{3}}{3}-
 \frac{8}{9}\cos^{4}\frac{\phi_{3}}{3},
\end{eqnarray}
where $\cos\phi_{3}=-p_{1}^{3}.$


\emph{c. Comprising the Ranges.}
 The relation
$\mathcal{I}_{2}[\vec{\kappa}]=\mathcal{I}_{2}[\vec{p}]$ leads to
$27\mathcal{K}=p^6_1$ and consequently $\phi_1=2\phi_2$. Therefore
we have
 \begin{eqnarray}\label{def11}
 \cos\frac{\phi_{1}+2\pi}{3}=2\cos^{2}\frac{\phi_{2}-2\pi}{3}-1.
\end{eqnarray}
\\Let $x=\cos\frac{\phi_{2}-2\pi}{3}$, one can obtain
\begin{eqnarray}\label{pkx}
&&p_{1}^{3}=\cos\phi_{2}=\cos(\phi_{2}-2\pi)=4x^{3}-3x \nonumber \\
&&\mathcal {K}=\frac{1}{27}p_{1}^{6}=\frac{1}{27}(4x^{3}-3x)^{2}.
\end{eqnarray}
Substituting them and Eq. (\ref{def11}) into Eqs. (\ref{def4}) and
(\ref{def7}), we get the first relation in Eq. (\ref{IpIk}),
$\underline{\mathcal{I}}_1[\vec{\kappa}]=\underline{\mathcal{I}}_1[\vec{p}]$.
In the same process, the angle $\phi_3=\pi-\phi=\pi-\phi_1/2$.
Setting $y=\cos (\phi_3 /3)$, we obtain
\begin{eqnarray}
p^3_1&=&-4y^3+3y, \nonumber \\
\mathcal{K}&=&\frac{1}{27}(4y^3-3y)^2.
\end{eqnarray}
These relations in company with Eqs. (\ref{def4}) and (\ref{def9})
lead to
$\overline{\mathcal{I}}_1[\vec{\kappa}]=\overline{\mathcal{I}}_1[\vec{p}]$,
which is the second relation in Eq. (\ref{IpIk}).



\end{document}